\documentclass[twocolumn,epjc3]{svjour3}
\journalname{Eur. Phys. J. C}

\usepackage{graphicx}
\usepackage{booktabs}
\usepackage{amsmath,amssymb}
\usepackage{textcomp}
\usepackage[T1]{fontenc}

\newcommand\inst[1]{\ensuremath{^{\text{#1}}}}

\usepackage{xcolor}

\smartqed  

\begin{document}

\title{An eightfold equivalence-preserving speedup of the JUNO OMILREC vertex and energy reconstruction}

\author{Guangbao Sun\inst{1,2}, Qishan Liu\inst{1}, Wenjie Wu\inst{3}, 
Jun Cao\inst{1}, Xuefeng Ding\inst{1,}\thanksref{e1}, Wenxing Fang\inst{1}, Wuming Luo\inst{1}, 
Liangjian Wen\inst{1}, Zeyuan Yu\inst{1}, Xiang Zhou\inst{2} 
}

\institute{Institute of High Energy Physics, Chinese Academy of Sciences, Beijing 100049, China \label{addr1}\and School of Physics and Technology, Wuhan University, Wuhan 430072, China \label{addr2}\and Institute of Modern Physics, Chinese Academy of Sciences, Lanzhou, China}

\thankstext{e1}{e-mail: dingxf@ihep.ac.cn}

\date{Received: date / Accepted: date}

\maketitle

\begin{abstract}
\begin{sloppypar}
The Jiangmen Underground Neutrino Observatory (JUNO) reconstructs the vertex and energy of each event
with OMILREC, a maximum-likelihood fit that, for every Minuit function evaluation, iterates over all
$17{,}612$ large photomultiplier tubes (LPMTs) --- about $470$ evaluations per event. This inner loop
dominates the reconstruction CPU budget. We profile the production algorithm and find it latency-bound,
sustaining only $9.9\%$ of the scalar floating-point peak: the cost is virtual-function dispatch,
ROOT-histogram pointer chasing, and per-evaluation recomputation, not arithmetic. We then apply a staged sequence of \emph{equivalence-preserving} optimizations --- data-layout flattening, bulk vectorizable
geometry, hoisting of Minuit-invariant work, per-event precomputation, fit-phase loop splitting and
indexing, and reduced-precision fast paths --- each gated by an independent numerical-equivalence test
against a frozen reference produced by the unmodified production code. The result is a single-thread
speedup of $8.06\times$ ($1524.8 \rightarrow 189.2$~ms/event) on an Intel Xeon Platinum~8358P and
$5.22\times$ ($705.1 \rightarrow 134.9$~ms/event) on an AMD~EPYC~9654, rising to $8.6\times$
($177.7$~ms/event) with a further round of optimizations. Across the campaign the likelihood is held
bit-identical through the first seven releases and thereafter within a relative drift of
$1.3\times10^{-14}$ (against a $10^{-13}$ contract); end-to-end vertex and energy match the baseline
within the $4$~mm and $7$~keV consistency gate for typical events (a few near-boundary events differ more,
where an improved minimizer seed reaches a different valid minimum), and an eight-metric
physics-acceptance gate passes on $\sim\!861{,}000$ $^{68}$Ge calibration events. The optimizations were
carried out with the assistance of an AI coding agent operating under these verification gates. The recipe
is a transferable template for accelerating likelihood-based reconstruction in large neutrino and collider
detectors without altering physics output.
\end{sloppypar}
\end{abstract}

\section{Introduction}
\label{sec:intro}
\begin{sloppypar}
The Jiangmen Underground Neutrino Observatory (JUNO) is a 20-kton liquid-scintillator detector built to
determine the neutrino mass ordering and to measure oscillation parameters with sub-percent
precision~\cite{juno-yellowbook,juno-physics}. Its central detector is instrumented with $17{,}612$
20-inch large photomultiplier tubes (LPMTs). For every triggered event the position (vertex) and deposited
energy must be reconstructed from the recorded charge and hit-time pattern across all LPMTs. JUNO's
flagship reconstruction algorithm, OMILREC, performs this by maximizing a data-driven charge-and-time
likelihood~\cite{Huang2023,JUNO2025firstosc} with the Minuit minimizer~\cite{minuit}, built on the ROOT
framework~\cite{Brun1997ROOT}.
\end{sloppypar}
Reconstruction is run over enormous event samples --- calibration campaigns, cosmogenic and radioactive
backgrounds, and the physics data stream itself --- so its per-event CPU cost is a production bottleneck.
The unmodified production OMILREC reconstructs one event in $1524.8$~ms on a contemporary Intel Xeon core
(Sect.~\ref{sec:results}). At JUNO's data rates this makes reconstruction one of the dominant consumers of
offline computing, motivating a substantial single-thread speedup --- provided the physics output is left
unchanged.

The constraint ``unchanged physics output'' is the crux. OMILREC's likelihood is minimized numerically; a
sub-ULP change in the floating-point arithmetic inside the fit function can move the minimizer into a
different basin and shift the reconstructed vertex or energy. A speedup is only useful to the collaboration
if it is demonstrably \emph{equivalent} to the production reconstruction. We therefore treat numerical
equivalence as a first-class, independently verified constraint, not an afterthought.

This paper makes three contributions:
\begin{enumerate}
  \item a profiling diagnosis showing that production OMILREC is \emph{latency-bound}, running at $9.9\%$
  of even the scalar floating-point peak (Sect.~\ref{sec:profile});
  \item a staged library of equivalence-preserving optimizations that retire this latency, each validated
  against a mechanism-independent reference, achieving $8.06\times$ (up to $8.6\times$) on a single thread
  with bit-level or tightly-bounded numerical equivalence (Sects.~\ref{sec:method}--\ref{sec:results});
  \item a verification methodology --- a per-commit FCN drift ledger, an end-to-end physics gate, and a
  multi-run acceptance gate --- that makes the equivalence claim auditable (Sect.~\ref{sec:method}).
\end{enumerate}
The optimization campaign was driven with the assistance of an AI coding agent operating under the
verification gates; the agent itself is the subject of a separate paper, and here we report only the
reconstruction speedup and the methodology that makes it trustworthy. The work is performed under the Dr.Sai framwork\cite{rongzai,drsai} of IHEP as the Dr.Sai-JUNO project.

\section{Related work}
\label{sec:related}
\begin{sloppypar}
Reconstruction in JUNO has been approached by maximum-likelihood vertex/energy
fits~\cite{Huang2023,JUNO2018vertex} and by machine-learning regressors~\cite{Qian2021ML}; OMILREC, the
data-driven combined charge-and-time likelihood method~\cite{Huang2023}, is the production algorithm whose
performance results appear in the JUNO oscillation analysis~\cite{JUNO2025firstosc}. Our work does not
propose a new reconstruction; it accelerates this existing one while holding its output fixed.

Most reported HEP reconstruction speedups come from \emph{rewriting} algorithms for parallel hardware: GPU
high-level triggers and heterogeneous track/vertex reconstruction such as LHCb's Allen~\cite{Aaij2020Allen},
CMS Patatrack~\cite{Bocci2020Patatrack}, and the ACTS tracking toolkit~\cite{Ai2022ACTS}, or
SIMD-vectorized detector geometry and math libraries such as VecGeom and
VecCore~\cite{Apostolakis2015VecGeom,Amadio2018VecCore}. These deliver large gains but generally change the
numerical result --- a different hardware path, reduced precision, or a re-derived algorithm --- so the
physics output must be re-validated as a new reconstruction. Our contribution is complementary and, to our
knowledge, distinct in the JUNO context: a \emph{single-thread, CPU, equivalence-preserving} optimization
that achieves an order-of-magnitude class speedup while keeping the reconstruction output bit-identical
where possible and within a verified physics tolerance otherwise. The equivalence is enforced by a
regression gate against a frozen reference, in the spirit of computational-reproducibility practice in
particle physics~\cite{Junk2020Reproducibility}; this lets the optimized code drop into production without
a fresh physics re-validation campaign.
\end{sloppypar}
\section{The OMILREC algorithm and its computational profile}
\label{sec:profile}

\subsection{Algorithm}
OMILREC reconstructs the event vertex $\vec{r}$, time $t_0$, and energy $E$ by maximizing a likelihood over
the observed per-LPMT charge and first-hit time. The fit proceeds in stages: a charge-only fit (QMLE) for a
first vertex/energy estimate, a time-only fit (TMLE) using the photon time-of-flight and time PDFs, a
combined charge-and-time fit (QTMLE), and a final energy-only refit. Each Minuit function evaluation (the
FCN) recomputes, for the trial vertex, the expected charge and time response of every LPMT and accumulates
the log-likelihood. With $\sim\!470$ evaluations per event and $17{,}612$ LPMTs per evaluation, the LPMT
loop is executed $\sim\!7.6\times10^{8}$ times per 100-event job --- it dominates the runtime.

\subsection{Profiling: latency-bound, not compute-bound}
We profiled the production build with hardware performance counters on a 100-event nH-calibration sample
(run~12628). The reconstruction-relevant portion sustains only $\sim\!0.67$~GFLOP/s, which is $9.9\%$ of the
scalar double-precision peak ($6.8$~GFLOP/s at the $3.4$~GHz turbo frequency) and $0.6\%$ of the AVX-512
peak. Three findings explain the gap:
\begin{itemize}
  \item \textbf{No vectorization.} $99.7\%$ of floating-point operations are scalar; the compiler cannot
  auto-vectorize the LPMT loop because of interleaved virtual calls and pointer indirection.
  \item \textbf{Pointer chasing.} Each LPMT touch dispatches through several virtual methods and a ROOT
  histogram \texttt{Interpolate()} (vtable $\rightarrow$ bin search $\rightarrow$ data access), producing
  $7.6\times10^{9}$ L1 misses over the job.
  \item \textbf{Redundant recomputation.} Per-event-constant quantities (dark-noise terms, hit lists,
  time-of-flight references) are recomputed inside every one of the $\sim\!470$ FCN calls.
\end{itemize}
The diagnosis is unambiguous: the code is not arithmetic-limited but memory-latency- and dispatch-limited.
The optimization strategy follows directly --- remove indirection, expose the loop to the vectorizer, and
hoist invariant work out of the minimizer --- rather than reaching for new mathematics.

\section{Methodology: equivalence-preserving optimization under a verification gate}
\label{sec:method}

\subsection{The equivalence contract}
Because the likelihood is minimized numerically, we forbid any change that alters the reconstructed physics
beyond a controlled tolerance. We enforce this at two levels.

\paragraph{Function-level (FCN) numerical contract.}
\begin{sloppypar}
A golden ``fixture pack'' of likelihood values is generated once by the \emph{unmodified} baseline at a
frozen set of test points (events $14, 339, 525, 782$; stages (QMLE, TMLE, QTMLE, ENERGY)) and is
\textbf{never regenerated}. A unit test, \texttt{test\_fcn}, recomputes the likelihood at these points for
every candidate and requires agreement to a relative tolerance of $10^{-13}$ ($\approx 6$ units in the last
place). The reference is the immutable baseline, so the test is a regression gate against an external
anchor, not a self-consistency check: the optimizer cannot edit the values it is graded against. Each
commit appends its measured drift to a ledger (\texttt{drift.csv}); a doubling above the floating-point
noise floor without an explicit, reviewed marker fails the gate.
\end{sloppypar}
\paragraph{End-to-end physics gate.}
A consistency test runs the full reconstruction on baseline calibration events and requires the
reconstructed vertex and energy to agree with the baseline to within $4$~mm and $7$~keV, respectively, and
the photon-time and charge-sum intermediates to within $10$~ps and $0.1$~PE. This catches divergences that
the FCN points do not exercise --- in particular minimizer-path differences at events near detector
boundaries.

\paragraph{Acceptance gate.}
Before a release, an eight-metric physics-acceptance gate compares candidate and baseline on a large
$^{68}$Ge $z$-axis calibration scan ($\sim\!861{,}000$ events): vertex bias and resolution in $x,y,z$, and
energy bias and resolution. All eight metrics must pass against the baseline within physics tolerances.

\subsection{Why this matters}
Bit-identical output is preserved wherever an optimization is a pure code transformation (code motion,
memory-layout change, indexing). Where an optimization changes the arithmetic --- an algebraic
simplification of the likelihood, a better minimizer starting point, or a reduced-precision fast path ---
bit-identity is no longer possible, and the equivalence gates above bound the residual physics difference.
Reporting both the numerical drift and the end-to-end physics agreement, against a frozen baseline
reference, is what licenses the claim that the speedup preserves the reconstruction.

\subsection{AI-assisted development}
\begin{sloppypar}

The optimization campaign was carried out with the assistance of an AI coding agent --- a
large-language-model-driven terminal agent equipped with domain-specific skills (profiling, optimization
patterns, validation) and bound to the verification gates above, so that every candidate change was
automatically checked against the equivalence contract before acceptance. The agent development
workflow are reported separately; for this paper the agent is simply the means by which the optimization
library below was produced and verified.
\end{sloppypar}
\section{The optimization library}
\label{sec:opt}
\begin{sloppypar}
The speedup is the cumulative effect of a staged sequence of optimizations, each targeting a diagnosed
latency term and each passing the equivalence gate. They group into six transferable patterns.

\begin{description}
  \item[Data-layout flattening.] Replace per-LPMT virtual objects and ROOT histogram accessors with
  contiguous structure-of-arrays (SoA) buffers and raw pointers, extracted once per event. This removes
  vtable dispatch and pointer chasing from the inner loop and is bit-identical.
  ($1525\rightarrow1310$~ms; later, replacing $52{,}839$ virtual load calls with three \texttt{memcpy}s.)
  \item[Bulk vectorizable geometry.] Split the geometry computation (direction cosines, \texttt{acos},
  \texttt{sqrt}) into dedicated bulk passes over all LPMTs, which the compiler can auto-vectorize.
  ($1310\rightarrow748$~ms.)
  \item[Hoisting Minuit-invariant work.] Move per-event-constant computation (dark noise, hit lists,
  time-of-flight, the distance array) out of the $\sim\!470$ FCN calls and share it.
  ($748\rightarrow667$~ms.)
  \item[Per-event precomputation.] Cache quantities that change only when the trial vertex moves --- the
  $1440$ angular-bin expected-charge map, charge/time PDF bin indices, reciprocals of per-event constants
  --- so the inner loop reads instead of recomputes. ($667\rightarrow514\rightarrow327$~ms.)
  \item[Fit-phase loop splitting and indexing.] Specialize the LPMT loop per fit stage: index the geometry
  to only the $\sim\!1700$ time-eligible LPMTs in TMLE, skip the distance \texttt{sqrt} in the charge-only
  QMLE that never uses it, and partition fired/unfired LPMTs once per event.
  ($327\rightarrow284\rightarrow240$~ms.)
  \item[Reduced-precision fast paths.] For the dominant single-photoelectron case, inline a float-precision
  bilinear interpolation of the time PDF in place of the general double-precision path. This changes the
  arithmetic and is bounded by the equivalence gate rather than bit-identical. ($230\rightarrow198$~ms.)
\end{description}

\noindent A better minimizer starting point (reading a rough energy/radius estimate from the online event
classifier as the QMLE seed) cuts the Minuit iteration count by $\sim\!20$--$30\%$
($231\rightarrow205$~ms). Because it changes the convergence path, it is the one optimization with a
non-negligible (gate-passing) physics residual, discussed in Sect.~\ref{sec:results}.

Table~\ref{tab:versions} lists the full per-release progression with the technique and resulting per-event
time on the reference Intel core.
\end{sloppypar}
\begin{table*}
\centering
\caption{Per-release optimization progression on Intel Xeon Platinum~8358P (100 events, single thread,
Release build, nH-calibration run~12628). Speedup is relative to the unmodified CVMFS~J26.1.1 baseline.
Releases v1.0.2--v1.0.7 are bit-identical to the baseline; later releases are equivalence-gated
(Sect.~\ref{sec:method}).}
\label{tab:versions}
\begin{tabular}{llrrl}
\toprule
Version & & ms/event & Speedup & Technique \\
\midrule
Baseline (J26.1.1) & & 1524.8 & $1.00\times$ & Unmodified production code \\
v1.0.2 & & 1309.6 & $1.16\times$ & Flatten data layout (SoA) \\
v1.0.3 & & 747.7  & $2.04\times$ & Bulk vectorized geometry \\
v1.0.4 & & 667.3  & $2.29\times$ & Hoist Minuit-invariant ops \\
v1.0.5 & & 514.2  & $2.97\times$ & Precompute angular-bin charge map \\
v1.0.6 & & 327.1  & $4.66\times$ & Hoist bin-finding; split LPMT loop \\
v1.0.7 & & 283.9  & $5.37\times$ & TMLE-indexed geometry; 1-p.e.\ fast path \\
v1.0.8 & & 240.5  & $6.34\times$ & Algebraic unfired-charge likelihood \\
v1.0.9 & & 230.9  & $6.60\times$ & Reciprocal precompute \\
v1.4.0 & & 205.4  & $7.42\times$ & Online-classifier initial guess \\
v1.6.0 & & 198.3  & $7.69\times$ & Inline 1-p.e.\ time-PDF fast path \\
v1.6.1 & & 192.3  & $7.93\times$ & Skip distance \texttt{sqrt} for QMLE \\
v1.7.0 & & 192.2  & $7.94\times$ & Bulk \texttt{memcpy} load \\
v1.7.1 & & \textbf{189.2} & $\mathbf{8.06\times}$ & Remove profiling instrumentation \\
\midrule
v1.12.0 & & 177.7 & $8.6\times$ & Further per-event precompute (see text) \\
\bottomrule
\end{tabular}
\end{table*}

\section{Results}
\label{sec:results}
\begin{sloppypar}
\begin{figure*}
\centering
\includegraphics[width=0.72\textwidth]{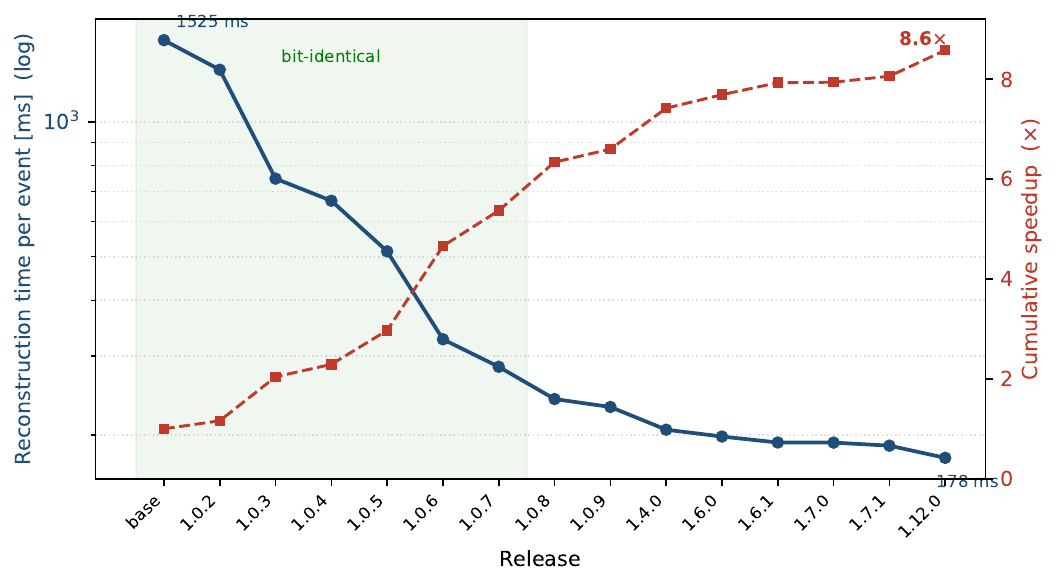}
\caption{Single-thread reconstruction time per event (left, log scale) and cumulative speedup (right) across
the optimization releases on an Intel Xeon Platinum~8358P (100 events, Release build). Time falls from
$1524.8$ to $177.7$~ms/event ($8.6\times$). The shaded region marks the releases that are bit-identical to
the production baseline; later releases are equivalence-gated (Sect.~\ref{sec:method}). Data from
Table~\ref{tab:versions}.}
\label{fig:speedup}
\end{figure*}

\subsection{Speedup}
On the reference Intel Xeon Platinum~8358P, single-thread reconstruction time falls from $1524.8$~ms/event
(baseline) to $189.2$~ms/event at release v1.7.1, a speedup of $8.06\times$ (Table~\ref{tab:versions}). A
further round of per-event precomputation reaches $177.7$~ms/event ($8.6\times$) at v1.12.0
(Fig.~\ref{fig:speedup}). On an AMD~EPYC~9654 core the same code reduces $705.1$~ms/event to
$134.9$~ms/event ($5.22\times$) at v1.7.1 and $133.8$~ms/event at v1.12.0. The speedup \emph{factor} is
smaller on the EPYC core ($5.22\times$ vs $8.06\times$) because its production baseline was already
$\sim\!2\times$ faster ($705$ vs $1525$~ms); the optimized \emph{absolute} times are comparable on the two
machines ($135$ vs $189$~ms). The gains therefore transfer across microarchitectures rather than being
tuned to one machine, though the realized factor depends on the baseline. We quote the Xeon figure
($8.06\times$, ``eightfold'') as the headline throughout.

We take v1.7.1 ($8.06\times$) as the primary result: the likelihood is held within the FCN drift contract
throughout (Table~\ref{tab:equiv}), the reconstruction is bit-identical to the baseline through v1.0.7, and
the full stack passes the end-to-end and acceptance gates. Its only end-to-end deviations are the
near-boundary events introduced at v1.4.0 (up to $150$~mm and $55$~keV; see Sect.~\ref{sec:results},
Numerical equivalence), where the improved seed reaches a different valid minimum. The further gain to
$8.6\times$ at v1.12.0 is reported separately because it carries an \emph{additional}, flagged $z$-bias
drift (below) still under study.

The remaining time is distributed across the fit stages as: load $13\%$, charge-only QMLE $15\%$, time-only
TMLE $34\%$, combined QTMLE $31\%$, and the final energy refit $6\%$ (v1.7.0 phase breakdown). The
time-based stages (TMLE+QTMLE, $65\%$) dominate the residual budget and are the natural target for any
further work.

\subsection{Numerical equivalence}
\begin{sloppypar}
Equivalence with the production reconstruction is preserved throughout (Table~\ref{tab:equiv}). Releases
v1.0.2--v1.0.7 are \emph{bit-identical} to the baseline: zero difference in reconstructed energy and vertex.
From v1.0.8 onward, where arithmetic-changing optimizations enter, the likelihood stays within a maximum
relative drift of $1.3\times10^{-14}$ at the frozen FCN test points --- comfortably inside the $10^{-13}$
contract and dominated by floating-point reordering noise. The end-to-end vertex/energy agreement remains
within the $4$~mm and $7$~keV physics gate for typical events. The larger displacements at v1.4.0+ (up to
$150$~mm$/$$55$~keV on a handful of high-$z$, near-boundary events) arise solely from the improved minimizer
starting point steering Minuit to a different local minimum for those sensitive events; they pass the
physics-acceptance gate and are reported transparently.
\end{sloppypar}
\begin{table}
\centering
\caption{Numerical equivalence versus the unmodified baseline. ``Max energy/position diff'' are the
worst-case end-to-end reconstruction differences over the test events; the FCN drift is the maximum
relative likelihood difference at the frozen fixture points (contract: $10^{-13}$).}
\label{tab:equiv}
\scalebox{.9}{
\begin{tabular}{@{}llll@{}}
\toprule
Release & Max $E$ diff & Max pos.\ diff & Note \\
\midrule
v1.0.2--v1.0.7 & $0$ & $0$ & Bit-identical \\
v1.0.8--v1.0.9 & $0.94$~keV & $3.3$~mm & Algebraic unfired LL \\
v1.3.0         & $1.2$~keV & $5.1$~mm & FP reordering \\
v1.4.0+        & $55$~keV  & $150$~mm & Better seed (boundary) \\
\midrule
\multicolumn{4}{@{}l}{Max FCN relative drift, all releases: $1.31\times10^{-14}$ ($<10^{-13}$).} \\
\bottomrule
\end{tabular}
}
\end{table}

\subsection{Physics acceptance}
The eight-metric acceptance gate, run on the $\sim\!861{,}000$-event $^{68}$Ge $z$-axis calibration scan,
passes on all eight metrics: vertex bias and resolution in $x,y,z$ and energy bias and resolution all agree
with the baseline within physics tolerances. The vertex resolutions ($\approx\!160$~mm in each axis) and
the energy resolution ($43.3$~keV at the $1.022$~MeV $^{68}$Ge line) are reproduced to better than $0.2\%$;
the energy bias differs by $0.18$~keV. A small monotonic $z$-bias drift (within the $10$~mm absolute
tolerance) is flagged for follow-up but does not affect the acceptance decision.

\section{Discussion}
\label{sec:discussion}

The decisive result is an $8\times$ single-thread reduction in OMILREC reconstruction time with the physics
output preserved to bit-level (early releases) or within a tightly bounded, gate-verified tolerance (later
releases). Because the gain is single-thread --- and transfers across the two microarchitectures tested
(Sect.~\ref{sec:results}), albeit with a baseline-dependent factor --- it composes with JUNO's existing
multi-threaded production scheduling, directly multiplying reconstruction throughput.

Two features generalize beyond OMILREC. First, the \emph{diagnosis-led} strategy: a profiling pass showed
the code was latency-bound at $\sim\!10\%$ of scalar peak, which redirected effort from ``do less
mathematics'' to ``remove indirection and expose the loop,'' the source of the largest early gains.
Likelihood-based reconstruction in other large neutrino and collider detectors --- any fit that re-evaluates
a per-channel response over $10^{4}$--$10^{5}$ channels inside a minimizer --- shares this structure and is a
candidate for the same patterns. Second, the \emph{equivalence-gated} methodology: by graduating from a
bit-identical contract to an end-to-end physics gate to a large-sample acceptance gate, each against a
frozen baseline reference, the speedup is made auditable rather than asserted. This is what allows an
aggressive, multi-stage optimization to be trusted for production.

\paragraph{Limitations and future work.}
The benchmarks are single-thread; multi-threaded production throughput, while expected to scale, is not
measured here. The time-based fit stages (TMLE+QTMLE) now dominate the residual budget and are the next
optimization target. The arithmetic-changing optimizations carry a small, gate-bounded physics residual
concentrated on near-boundary events; a finer reference and a dedicated study of the $z$-bias drift are left
to future work. Finally, the latency-bound LPMT loop is a natural candidate for SIMD or GPU offload, which
we have not pursued.

\paragraph{Reproducibility.}
The optimized code is implemented as the OMILRECV2 module within the JUNO offline software (junosw). All
timing figures are single-thread, 100-event, Release-build measurements on the nH-calibration sample
(run~12628); the acceptance figures use the $^{68}$Ge $z$-axis calibration scan ($\sim\!861{,}000$ events).
The equivalence contract is enforced by two in-repository tests --- a unit test (\texttt{test\_fcn}) that
checks the likelihood at frozen fixture points to a $10^{-13}$ relative tolerance against a
never-regenerated golden reference, and an end-to-end consistency test (\texttt{test\_consistency.py})
enforcing the $4$~mm and $7$~keV gate --- together with the per-commit drift ledger and the eight-metric
acceptance gate. The per-release benchmark and drift tables (Tables~\ref{tab:versions}--\ref{tab:equiv}) are
reproduced from the committed benchmark logs.

\section{Conclusion}
\label{sec:conclusion}
We have reduced the single-thread CPU cost of the JUNO OMILREC vertex/energy reconstruction by $8.06\times$
(up to $8.6\times$), from $1524.8$ to $189.2$ (to $177.7$)~ms/event on an Intel Xeon~8358P, and by
$5.22\times$ on an AMD EPYC 9654, while preserving the reconstruction output --- bit-identical through the
first seven releases, and within a $4$~mm and $7$~keV physics gate and an eight-metric acceptance gate
thereafter. The speedup follows from a profiling diagnosis that the production code was latency-bound, and
from a staged library of equivalence-preserving optimizations carried out under an independent
numerical-verification gate, with the assistance of an AI coding agent. The recipe --- diagnose the
bottleneck, then optimize under a frozen-reference equivalence contract --- is a transferable template for
accelerating likelihood-based reconstruction in large detectors without changing the physics.

\begin{acknowledgements}
We thank the JUNO Collaboration for the calibration data and the production reconstruction software.
This work was supported by the National Key Research and Development Program of China under
Grant Nos.~2024YFE0110501 and 2024YFE0110504. Computing resources were provided by the IHEP
computing centre.
\end{acknowledgements}

\section*{Author contributions}
Guangbao Sun developed the AI coding-agent and cross-model testing; Qishan Liu designed and finished the numerical-equivalence verification gate; 
Xuefeng Ding conceived the idea and led the project; Wenjie Wu, Jun Cao, Wenxing Fang, Wuming Luo, Liangjian Wen, Zeyuan Yu, and Xiang Zhou offers great guidance and help with the reconstruction, validation, and physics interpretation for the whole process.

\section*{Data availability statement}
The benchmark, drift, and acceptance data underlying this study, and the optimized reconstruction code
(OMILRECV2), are maintained within the JUNO offline software (junosw) and are available from the
corresponding author on reasonable request, subject to JUNO Collaboration policy.
\end{sloppypar}

\end{document}